# Electro-optic frequency comb-empowered precise measurement of the dynamic frequency of a laser

Weiwei Yang [1†], Xingyu Jia[1†], Jingyi Wang[2], Xinlun Cai[2*], Yang Li[2*], Guanhao Wu[1*]

[1] State Key Laboratory of Precision Measurement Technology and Instruments, Department of Precision Instrument, Tsinghua University; Beijing, 100084, China.

[2] State Key Laboratory of Optoelectronic Materials and Technologies, School of Electronics and Information Technology, Sun Yat-sen University; Guangzhou, 510275, China.

*Corresponding author. Email: caixlun5@mail.sysu.edu.cn; liyang328@mail.sysu.edu.cn; guanhaowu@mail.tsinghua.edu.cn

† These authors contributed equally to this work.

**Abstract:** Frequency-modulated lasers (FMLs) are widely used in spectroscopy, biology, and LiDAR. These applications' performance highly depends on the fast and precise tracking of the FMLs' absolute frequencies, which remains a challenge. Here we demonstrated integrated lithium niobate electro-optic frequency combs with arbitrarily tunable repetition rates and 29.45-nm bandwidth, enabling precise tracking of the absolute frequency of FML with a chirp rate as high as $2\times10^{18}$ Hz/s, which is over 3 orders of magnitude beyond the state-of-the-art. This method enabled frequency-modulated continuous-wave ranging using an FML with severe mode hops, unlocking great potential for improving the ranging resolution and acquisition rate. Our method built the foundation of FML-based high-precision measurement of frequency, distance, and time, leading to profound implications in fundamental science and engineering applications.

## Introduction

As one of the most widely used light sources for metrology, frequency-modulated lasers (FMLs) — continuous-wave (CW) lasers whose central frequencies can be arbitrarily tuned — are the essential devices of tunable diode laser absorption spectroscopy (TDLAS) [1–3], frequency-modulated continuous-wave (FMCW) LiDAR[4–6], and optical coherence tomography (OCT) [7–9]. The precision and acquisition rate of these applications are highly dependent on the precise and fast tracking of FMLs' instantaneous absolute frequencies[10].

Frequencies of FMLs are conventionally measured by cavities[11,12], interferometers[13], and frequency combs[14–23]. Cavity- and interferometer-based methods feature simple experiment setups and moderate precision. However, cavity- and interferometer-based methods trace the frequency of FML to cavities' lengths and interferometers' unbalanced lengths which are affected by dispersion and ambient environmental variation[12,13,24], limiting the accuracy and stability. On the other hand, frequency combs — a series of equally spaced longitudinal modes in the spectrum — obtain an FML's frequency by measuring the frequency difference between the frequency under test and its adjacent comb line's frequency $f_n = f_{ceo} + n \times f_r$ where $f_{ceo}$ is the carrier envelope offset frequency, $f_r$ is the repetition rate in-between neighboring comb lines, $n$ is the index number of comb lines. Because $f_{ceo}$ and $f_r$ can be referenced to the atomic clock[25], the frequency comb-based method can achieve high precision, accuracy, and stability. However, the frequency comb-based method's acquisition rate is limited by the low repetition rate of conventional frequency combs based on mode-locked lasers[14], preventing the measurement of the instantaneous absolute frequencies of FMLs with high chirp rates[26–28] (more details see Supplementary Text S1).

Furthermore, many applications require modulating an FML's frequency over a broad band at a high chirp rate to improve the performance. Particularly, an FMCW LiDAR always desires a higher ranging resolution and a higher acquisition rate, necessitating an FML with a wider modulation bandwidth and a higher chirp rate[29]. However, such an FML usually induces more mode hops[30,31], preventing the FMCW LiDAR from unambiguously retrieving the distance (Fig. 1B, right). So far, to the best of our knowledge, FMCW LiDAR based on a mode-hop FML has not been demonstrated.

Here we demonstrated high-precision, real-time measurement of FML's instantaneous absolute frequency by leveraging the high repetition rates of electro-optic (EO) frequency combs (Fig. 1). By designing integrated lithium niobate EO combs with arbitrarily tunable repetition rates and 29.45-nm bandwidth (Fig. 1A), we can actively adjust our dual EO comb-based method's acquisition rate by tuning the repetition rates. Then, by selecting a repetition rate of 10 GHz, we achieve an acquisition rate of 10 GHz, allowing the tracking of a high-speed linearly chirped signal with a chirp rate up to $2 \times 10^{18}$ Hz/s, which is over 3 orders of magnitude higher than the state-of-the-art fast absolute-frequency measurement method[14]. Furthermore, our dual EO comb-based method also enables measuring the absolute frequency of an arbitrary mode-hop FML, leading to the first FMCW ranging with an arbitrary mode-hop FML. Finally, we demonstrate 3D imaging at a distance of 20 m using a commercial mode-hop laser (Fig. 1B). The system covers a field of view of 24° × 20° and successfully reconstructs the shapes of a real building and a real person.

As shown in Fig. 1C, compared with the mode-locked laser-based frequency measurement method, our EO comb's higher repetition rate results in a much higher acquisition rate, enabling the measurement of frequency-agile FMLs. In comparison with the Kerr comb-based frequency measurement method, our EO comb's arbitrarily tunable repetition rate results in a much greater flexibility in tuning the acquisition rate, leading to compatibility with various data acquisition and processing hardware.

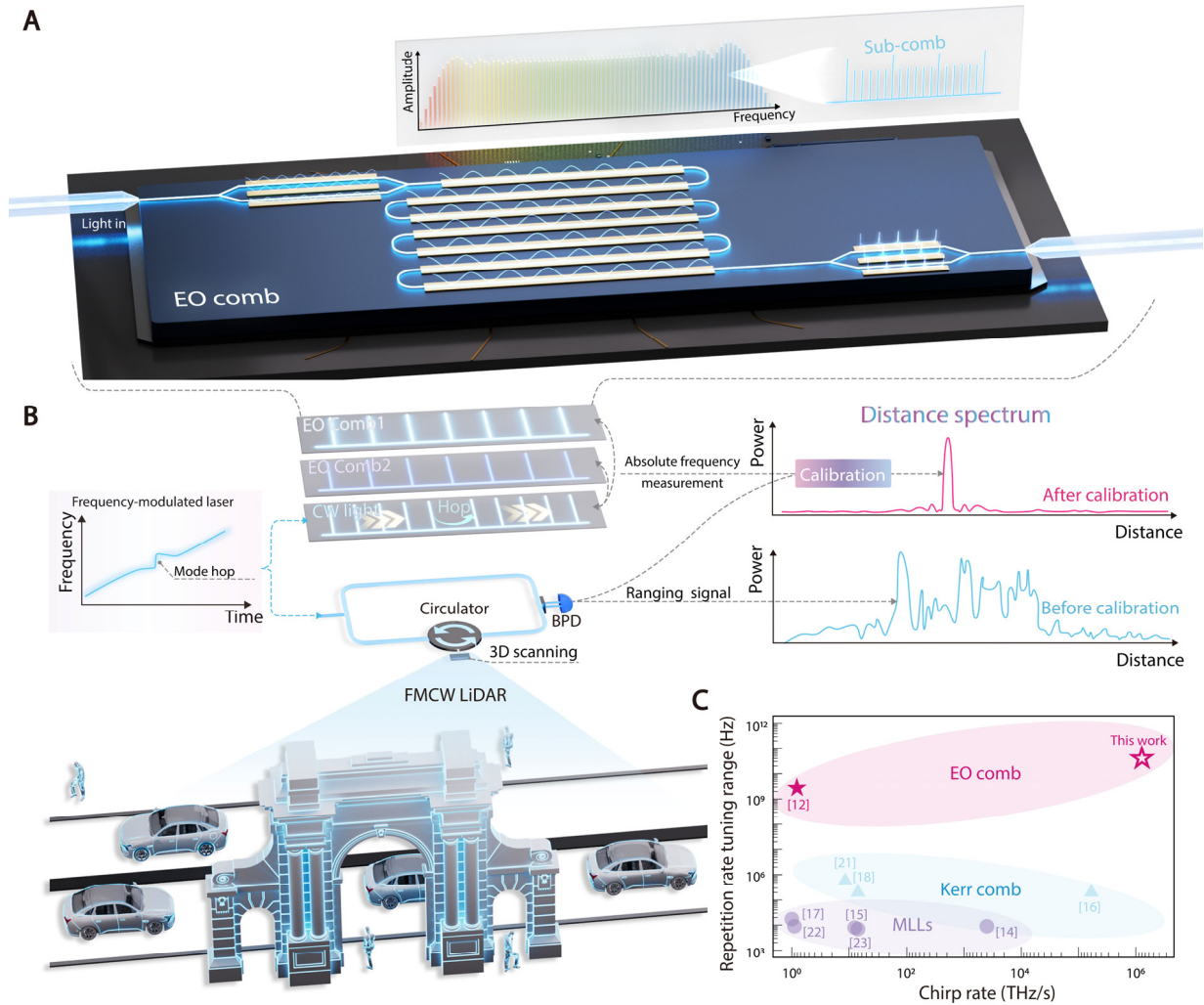

**Fig. 1. Principle of dual electro-optic (EO) comb-based frequency measurement and its application in the calibration of FMCW LiDAR.** (**A**) Schematic of integrated lithium niobate electro-optic frequency comb. The inset shows the EO comb, consisting of a high-repetition-rate main comb and a low-repetition-rate sub-comb. (**B**) Schematic of FMCW LiDAR based on a mode-hop frequency modulated laser. This LiDAR's ranging signal is calibrated by the absolute frequency measured by a dual EO comb-based absolute frequency measurement module, enabling the unambiguous distance retrieval. BPD: balanced photodetector. (**C**) Compared with other methods, our method can measure the frequency of an FML with a higher chirp rate. And, our EO comb's repetition rate can be adjusted over a broader range, leading to a wider tuning range of acquisition rate. Detailed references can be found in Supplementary Table S1. MLLs: mode-locked lasers.

**Device for FML frequency measurement**

We cascaded an amplitude modulator and three phase modulators to generate a broadband flat-top EO comb with an arbitrarily tunable repetition rate (Figure 2). To achieve an EO comb with an arbitrarily tunable repetition rate, we designed the device based on EO phase modulators without any optical and electric resonant structure, circumventing the limitations of the optical resonator's free spectral range and the electric resonator's resonant condition on the tuning range of the repetition rate. To increase the number of comb lines without increasing the driving voltage, we designed each phase modulator with a modulation length as long as 2.68 cm, featuring a half-wave voltage as low as 2 V. We developed a compactly packaged electro-optic frequency comb[32], as shown in the photograph (Fig. 2B). Furthermore, we drove all the amplitude and phase modulators with a high-frequency microwave signal, resulting in a high repetition rate $f_{r,1}$ of 25 GHz (Fig. 2A, the top-left inset). Thirdly, we cascaded three identical phase modulators (more details see Supplementary Text S2), finally leading to an EO comb with a bandwidth as high as 29.45 nm (Fig. 2C). To flatten this broadband comb, we fine-tuned the amplitude and phase of the driving microwave signals of the amplitude modulator and three phase modulators, as well as the DC bias of the amplitude modulator.

We cascaded an amplitude modulator after the phase modulators to reduce the repetition rate of the flat-top comb. To cover the tuning range of the FML under test, we usually generate the flat-top comb with a repetition rate as high as possible considering the limitations of our microwave equipment, resulting in a broadband flat-top comb with a high repetition rate. However, such a high repetition rate may be beyond the sampling rate of the oscilloscope's analog-to-digital converter. To reduce the repetition rate, we cascaded an amplitude modulator after the phase modulators[33]. Such an amplitude modulator is driven by a microwave frequency comb with a repetition rate $f_{r,2}$, which can be, in principle, an arbitrary frequency (Fig. 2A, the top-right inset). $f_{r,1}$ must be integer multiples of $f_{r,2}$, leading to a flat-top broadband EO comb with a repetition rate of $f_{r,2}$. For example, the spectra of these flat-top EO combs with $f_{r,2}$ of 5 GHz, 1 GHz, 500 MHz, and 250 MHz are shown in Figure 2D.

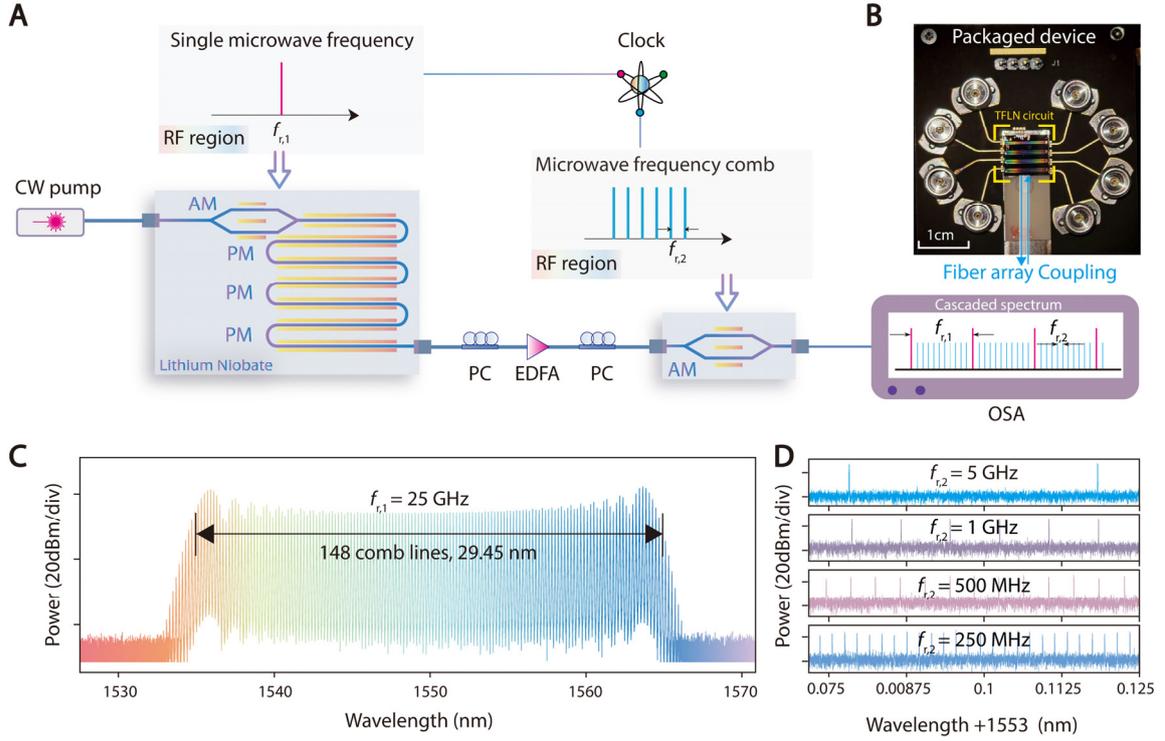

**Fig. 2. Integrated lithium niobate EO comb for measuring the dynamic frequency of FML.**
(**A**) Schematic of two-stage EO comb generation. CW: continuous wave; AM: amplitude modulator. PM: phase modulator. PC: polarization controller. RF radio frequency. EDFA: erbium-doped fiber amplifier. OSA: optical spectrum analyzer. (**B**) Device photograph. TFLN: thin film lithium niobate. (**C**) Measured spectrum of the first broadband EO comb with a high repetition rate. (**D**) Measured spectra of the final EO combs with repetition rates of 5 GHz, 1 GHz, 500 MHz, and 250 MHz.

## Absolute frequency measurement module

We measure the instantaneous absolute frequency $f_{\text{CW}}$ of an FML via its beat frequencies with two EO combs with slightly different repetition rates. As shown in Figure 3A (left), as $f_{\text{CW}}$ sweeps over the spectrum of EO comb 1, $f_{\text{CW}}$ beats with all the comb lines whose frequencies are $f_0 + n \times f_{r,2}$. To guarantee that $f_{\text{CW}}$ only beats with the most adjacent comb line at a given time, we set the oscilloscope's sampling rate equal to $f_{r,2}$, leading to the instantaneous beat frequency $f_{b,1}$ and the beat frequency map (red lines in Figure 3A, right). Similarly, $f_{\text{CW}}$ beats with EO comb 2, which shares the same pump CW laser as EO comb 1 but has a slightly different repetition rate $f_{r,2} + \delta f_r$, resulting in the instantaneous beat frequency $f_{b,2}$ and the beat frequency map (blue lines in Figure 3A, right).

We obtain the comb-line number based on the measured beat frequencies $f_{b,1}$ and $f_{b,2}$ via $n = (f_{b,1} - f_{b,2})/\delta f_r$, allowing us to calculate the real-time absolute frequency

$f_{\text{CW}} = f_0 + n \times f_{r,2} + f_{b,1}$. This process is explained by three representative times 1, 2, and 3. At time 1 (green dot in Figure 3A, left), $f_{\text{CW}}$ beats with the identical pump CWs of EO combs 1 and 2, leading to the same measured beat frequencies $f_{b,1} = f_{b,2}$ (green dots in Figure 3A, right). Hence, we can obtain the comb-line number as $n = (f_{b,1} - f_{b,2})/\delta f_r = 0$, allowing us to get the absolute frequency $f_{\text{CW}} = f_0 + 0 \times f_{r,2} + f_{b,1} = f_0 + f_{b,1}$. At time 2 (yellow dot in Figure 3A, left), $f_{\text{CW}}$ beats with the first comb lines of EO combs 1 and 2, resulting in distinct measured beat frequencies $f_{b,1} - f_{b,2} = \delta f_r$, which is manifested as the difference between the two yellow dots in the beat frequency map (Fig. 3A, right). So, we can obtain the comb-line number via $n = (f_{b,1} - f_{b,2})/\delta f_r = 1$, allowing us to get the absolute frequency $f_{\text{CW}} = f_0 + f_{r,2} + f_{b,1}$. At time 3 (red dot in Figure 3A, left), $f_{\text{CW}}$ beats with the second comb lines of EO combs 1 and 2, leading to different measured beat frequencies $f_{b,1} - f_{b,2} = 2\delta f_r$. This difference is depicted by the two red dots in the beat frequency map (Fig. 3A, right). So, we can calculate the comb-line number via $n = (f_{b,1} - f_{b,2})/\delta f_r = 2$, allowing us to obtain the absolute frequency $f_{\text{CW}} = f_0 + 2f_{r,2} + f_{b,1}$.

Figure 3B shows the experimental setup for measuring the real-time absolute frequency $f_{\text{CW}}$. We pump two integrated lithium niobate EO combs with a CW laser, generating EO comb 1 and EO comb 2. Then, to achieve complex beat frequency detection and real-time data processing, we used an optical in-phase/quadrature demodulator followed by a pair of balanced photodetectors to probe the beat frequencies between the $f_{\text{CW}}$ of FML under test and the EO combs.

We performed experiments to demonstrate our method's capability in measuring a mode-hop FML whose frequency shows the letters "THU" as a function of time. The generation of this FML is described in Supplementary Text S3. As shown in Figure 3C, the beat frequency between the FML and an EO comb gives the difference between the FML's instantaneous absolute frequency and a certain comb line, leading to ambiguities when the beats between distinct instantaneous absolute frequencies and their corresponding most adjacent comb lines give the same beat frequency — the four segments of beat frequencies near -0.23 GHz. Such ambiguities can be resolved by obtaining the comb-line numbers using our method, leading to the absolute frequencies in Figure 3E. This result shows that our method can distinguish the same beat frequencies near -0.23 GHz (Fig. 3C) into four segments of distinct absolute frequencies near 193.479 THz, 193.481 THz, and 193.483 THz.

We further verified that our method can precisely measure the instantaneous frequency of a frequency-agile FML. We measured a frequency-agile FML with a chirp rate as high as 2,000,000 THz/s. Figure 3D shows that our measured absolute frequency can faithfully resemble this high chirp rate, which is over 3 orders of magnitude higher than the state-of-the-art fast absolute-frequency measurement method[14]. Moreover, we performed experiments to evaluate our method's uncertainty, resolution, and adaptability to arbitrarily chirping and mode-hop behaviors (more details see Supplementary Text S4).

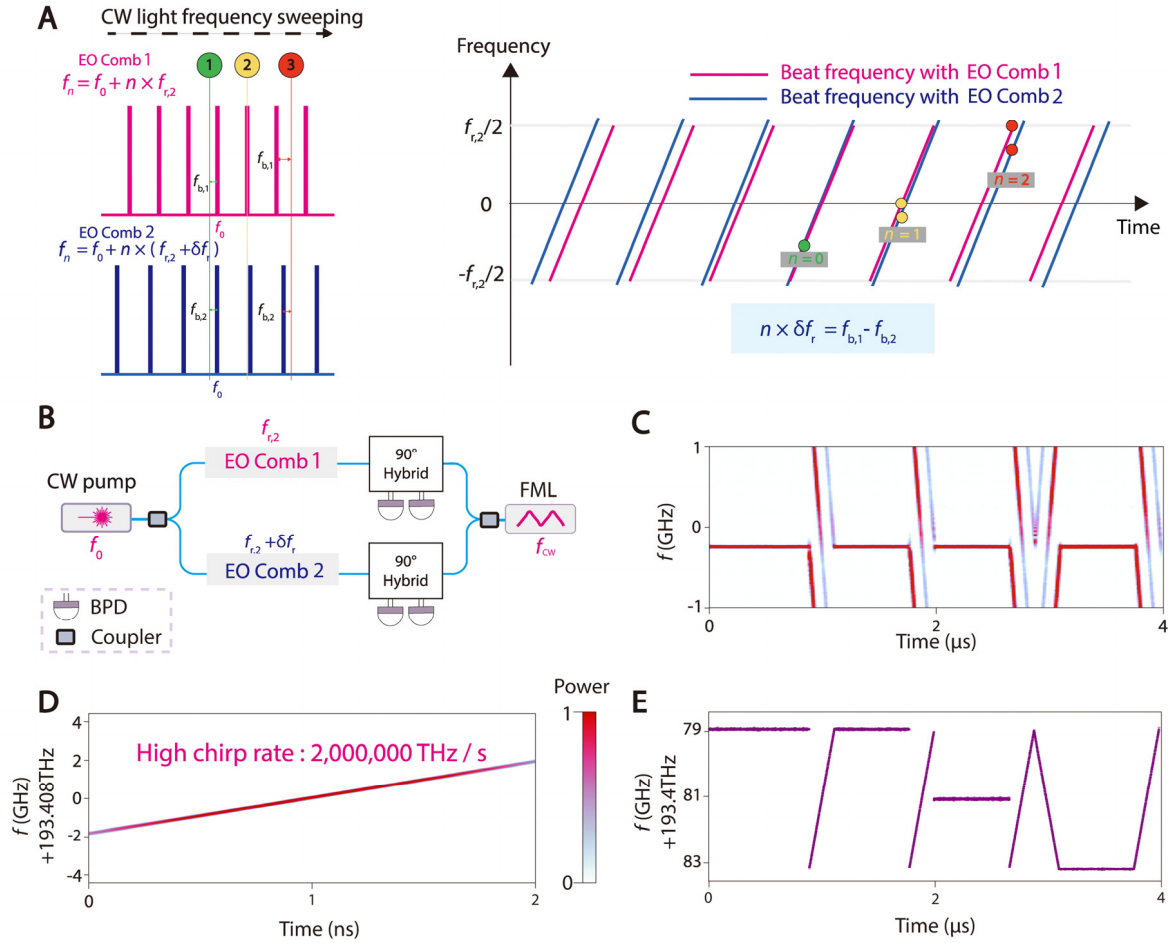

**Fig. 3. Principle, experimental setup, and measurement results of absolute frequencies of FMLs based on dual EO combs.** (**A**) Principle schematic and (**B**) experimental setup of dual EO comb-based measurement of a frequency modulated laser (FML). BPD: balanced photodetector. 90° hybrid: 90° optical hybrid module. (**C**) Frequency-time diagram of the beat frequency between a mode-hop FML and an EO comb. (**D**) The absolute frequency of a frequency-agile FML. (**E**) The absolute frequency corresponding to the beat frequency in (**C**).

**FMCW ranging with mode-hop FML**

It is imperative to precisely and fast trace the mode-hop FML's absolute frequency to enable FMCW ranging with sufficient ranging resolution and acquisition rate. To improve the FMCW's ranging resolution and acquisition rate, we need to increase the FML's modulation bandwidth over a shorter modulation period. However, an FML with a broader modulation bandwidth and/or a shorter modulation period usually induces more severe mode hops and nonlinearities (Fig. 4A, middle)[31]. These mode hops and nonlinearities result in multi-peaks (Fig. 4A, pink curve on the left) and broadband (Fig. 4A, blue curve on the right) behaviors, respectively, in the distance spectra, preventing us from unambiguously retrieving the distance. If we could precisely and fast calibrate these mode hops and nonlinearities, we should be able to obtain a single peak in the distance spectrum as that provided by a mode-hop-free linear-chirp FML (Fig. 4A, black curves in left and right).

To demonstrate the feasibility of using mode-hop FMLs in FMCW ranging, we performed FMCW experiments using an FML with severe mode hop and nonlinearity. As shown in Figure 4B, we generated a mode-hop and nonlinear FML by externally modulating a CW laser (more details see Supplementary Text S3). Then, we input this mode-hop FML into an FMCW setup consisting of an unbalanced Mach-Zehnder interferometer whose unbalanced length equals the distance under test. We also input such a mode-hop FML into an absolute frequency measurement module, synchronously outputting the instantaneous absolute frequency of the mode-hop FML.

We used our method to reconstruct the mode-hop FML-based FMCW signal, enabling the first FMCW ranging based on a mode-hop FML. Based on the measured time-domain waveform of the FMCW signal (Fig. 4C), we used our method to measure this signal's absolute frequency, showing severe mode hop, nonlinearity, and frequency overlapping (Fig. 4E). According to this absolute frequency, we reconstructed the FMCW signal in the spectral domain (Fig. 4D) by resampling the time-domain waveform in Figure 4C. By transforming the original and reconstructed FMCW signals into the distance domain, we obtained the FMCW's ranging results (Fig. 4F). To handle the non-uniformly frequency-sampled signals, we employed the Lomb–Scargle periodogram[34] (more details see Supplementary Text S5). Due to the errors introduced by mode hop, nonlinearity, and frequency overlapping, the original FMCW result shows several peaks over a wide distance range, preventing the effective retrieval of the distance under measurement. In contrast, the reconstructed FMCW result depicts a sharp and unique peak, leading to the unambiguous retrieval of the distance under measurement!

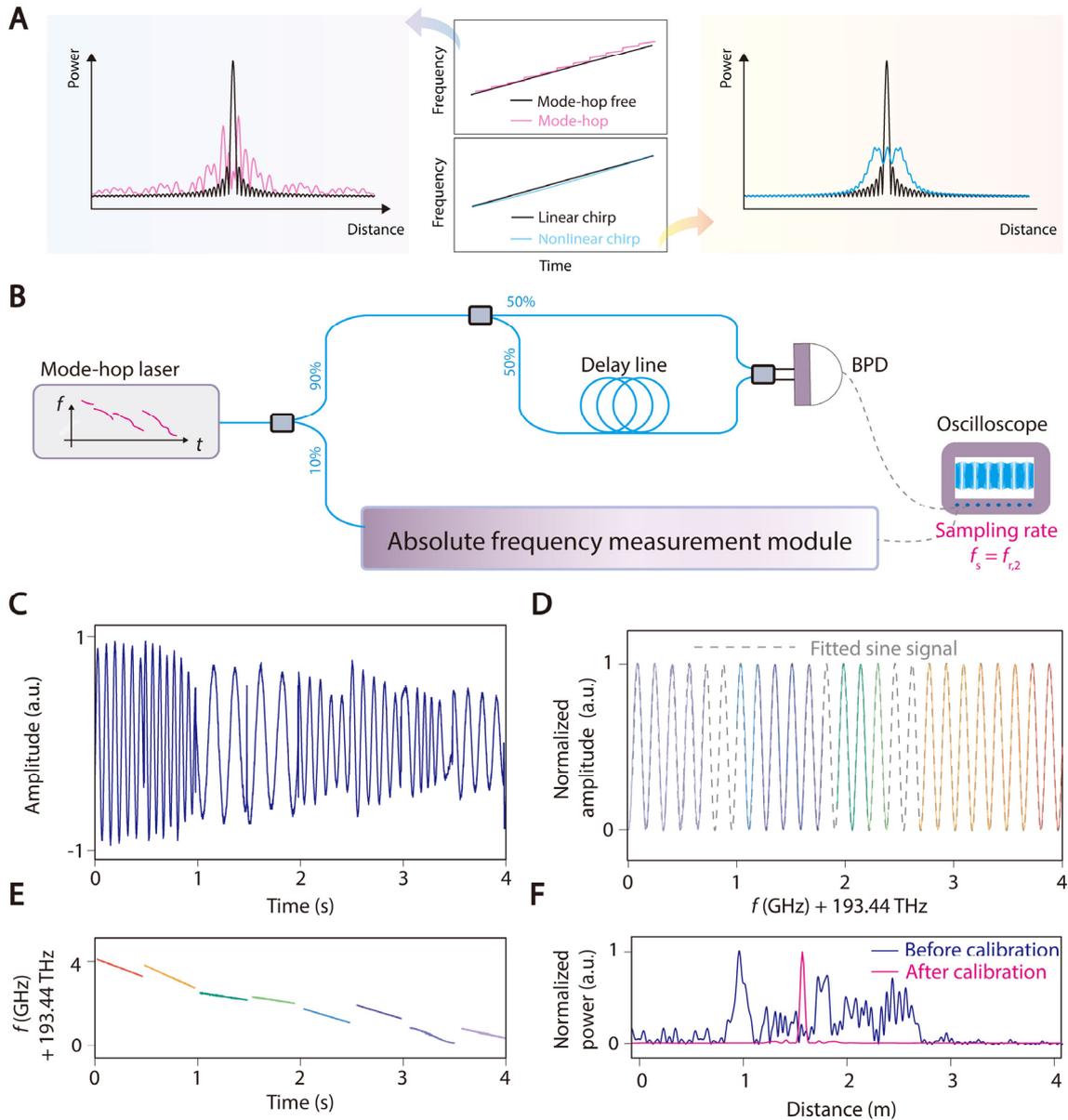

**Fig. 4. The experimental setup, schematic, and measurement results of FMCW using our method.** (**A**) Simulated distance spectra of FMCWs based on mode-hop-free, mode-hop, linear-chirp, and nonlinear-chirp FMLs. (**B**) Experimental setup. BPD: balanced photodetector. (**C**) Measured time-domain waveform of FMCW signal with mode hops in (**E**). (**D**) Measured FMCW signal reconstructed based on the waveform in (**C**) using the absolute frequency in (**E**). (**E**) Measured absolute frequency obtained using our method. (**F**) Mode-hop FML-based FMCW's ranging results before and after the calibration using our method.

### 3D FMCW LiDAR using a mode-hop FML

To demonstrate our method's capability in achieving a 3D FMCW LiDAR using a commercial mode-hop FML, we performed 3D imaging by combining our method and an FMCW LiDAR based on a mode-hop frequency-modulated DFB laser. The frequency sweep has a period of 1 ms and a bandwidth of approximately 25 GHz, with an unknown mode hop occurring in the middle of the sweep (Supplementary Text S6). As shown in Figure 5A, we used a circulator to separate the transmission and reflection measurement beams. Then, we employed a collimator to transform the guided light in the fiber to the propagation light in free space, focusing the beam on the target objects at a stand-off of 20 m. Finally, we used two galvos to scan the beam along two orthogonal directions, leading to the 3D imaging of the target objects. To precisely measure the distance for each pixel, we used our method to measure the absolute frequency of the mode-hop frequency-modulated DFB laser used in the FMCW LiDAR. Then, this measured absolute frequency was used to reconstruct the FMCW result. In this absolute-frequency measurement process, to avoid an ultrahigh acquisition rate and the corresponding huge memory, we selected a 200-MHz acquisition rate. In this way, we generated high-precision point clouds of diffusely scattering target objects including the historic old gate of the Tsinghua University and a real person (Figs. 5B-5D). More details can be found in Supplementary Movies S1 & S2.

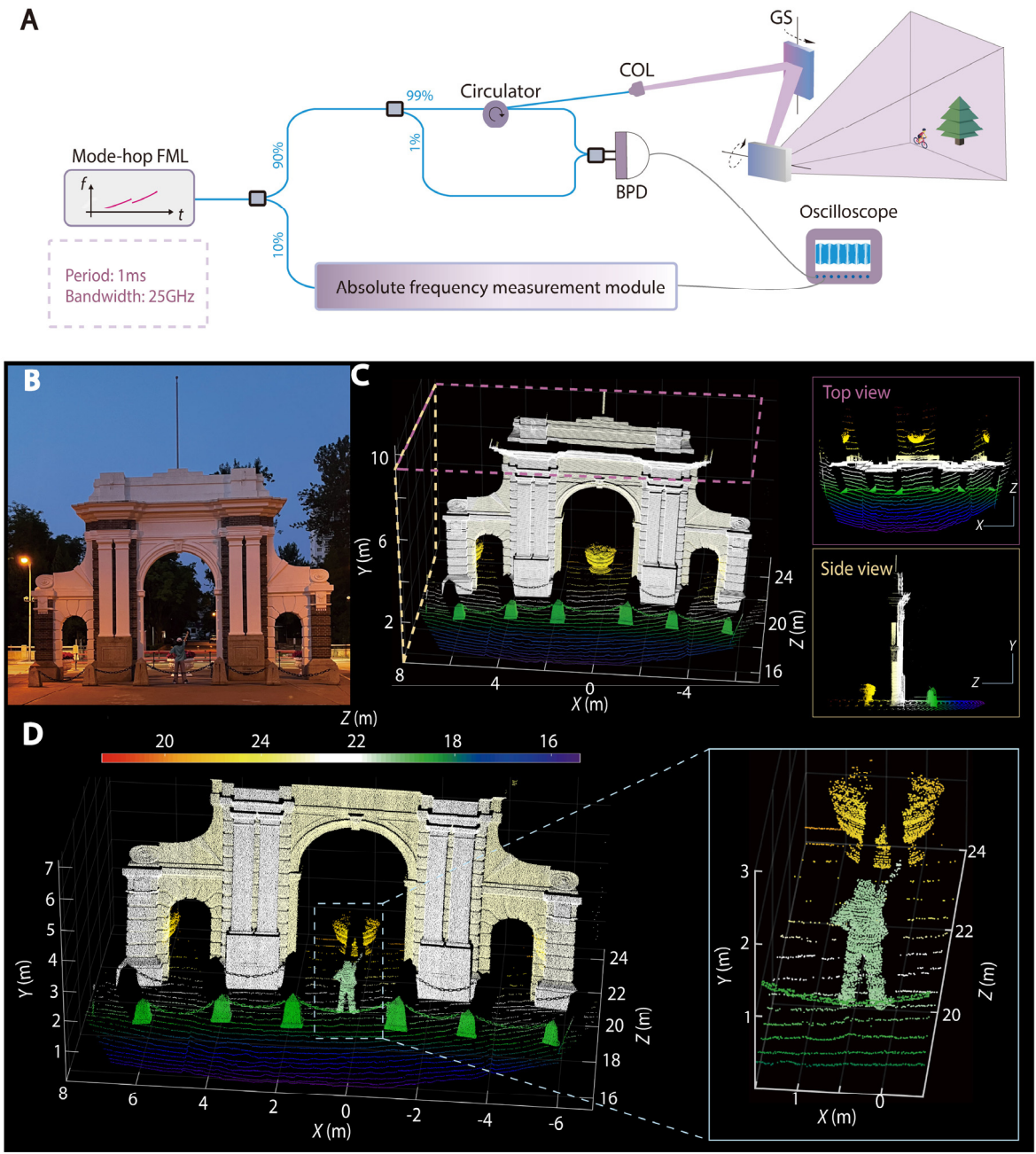

**Fig. 5. 3D FMCW LiDAR using a mode-hop frequency-modulated distributed feedback (DFB) laser. (A)** Experimental setup. COL: collimator. BPD: balanced photodetector. GS: Galvo scanning system. **(B)** Photograph of the target objects. **(C)** Measured point cloud of the Old Gate of Tsinghua Garden (4000×500 points), shown from a 3D perspective, top view, and side view. **(D)** Measured point cloud of a real person in front of the Old Gate of Tsinghua Garden (1000×200 points), along with a zoomed-in view of the person.

**Conclusion and discussion**

We achieved an EO comb with an arbitrarily tunable repetition rate and a bandwidth up to 29.45 nm by cascading several phase modulators and amplitude modulators. Based on dual EO combs, we proposed and demonstrated an absolute frequency measurement method, leading to the precise measurement of the instantaneous frequency of a frequency-agile FML with a chirp rate as high as 2,000,000 THz/s. Using this absolute frequency measurement method, we measured the absolute frequency of an FML with severe mode hop, nonlinearity, and frequency overlapping, leading to the first FMCW distance measurement with a mode-hop FML. By combining our method with a 3D imaging FMCW LiDAR based on a mode-hop frequency-modulated DFB laser, we generated high-precision point clouds of diffusely scattering target objects with a stand-off of 20 m. Beyond LiDAR, our precise absolute frequency measurement method also has broad applications in various areas such as spectroscopy, precise microwave and terahertz frequency measurement, spatial-temporal standard, and biomedical science and engineering.

We envision our work's future development in the following aspects. Firstly, by monolithically integrating a semiconductor CW laser[35–39] and silicon photodetectors[37,40,41] on a single thin-film lithium niobate chip, we could achieve an integrated absolute frequency measurement module. Secondly, our absolute frequency measurement method can be extended to other telecom wavelengths from S to U bands by shifting the EO comb's pump wavelength[32]. Thirdly, our method provides an effective way to compensate for one of the biggest limitations — mode hop — of broadband tunable CW lasers[42–46]. For example, using our method, we could calibrate the mode hops of broadband tunable CW lasers of FMCW LiDARs, significantly improving the ranging resolution.

## Materials and methods

### *Fabrication process*

The integrated electro-optic device (Fig. 2B) was fabricated on a 6-inch x-cut thin-film lithium niobate (TFLN) on insulator wafer using a standard planar process. The wafer comprised a 360-nm-thick LN membrane, a 9-µm-thick buried $SiO_2$ layer, and a 675-µm-thick high-resistivity silicon substrate. Passive photonic components — including waveguides, grating couplers, multimode interference (MMI) couplers, and edge couplers — were patterned using deep ultraviolet (DUV) lithography and etched 180 nm into the LN layer by inductively coupled plasma (ICP) etching. A 1.2-µm-thick $SiO_2$ cladding was deposited via plasma-enhanced chemical vapor deposition (PECVD) to minimize metal electrode-induced optical absorption. Gold RF electrodes with a thickness of 900 nm were then formed using e-beam evaporation and a lift-off process. A 500-nm $SiO_2$ top cladding was subsequently deposited for surface protection. Finally, to improve velocity matching and reduce microwave dielectric loss, 25-µm silicon substrate underlying the main microwave field was removed by dry etching.

The packaged electro-optic frequency comb generator was designed for robust and efficient system integration. A commercial fiber array was first aligned to the edge couplers of the photonic integrated circuit (PIC) to enable low coupling-loss optical input and output. The chip, equipped with fiber pigtails, was then embedded into a recessed cavity on a customized evaluation printed circuit board (PCB) to facilitate radio frequency (RF) and direct current (DC) wire bonding. Gold wire bonds connected the on-chip pads to corresponding metal pads on the PCB, which were routed through coplanar waveguides to 1.85 mm coaxial RF connectors for microwave signal input and termination. For the amplitude modulator, which operated under low RF power, an on-chip 50-Ω terminating resistor was used without inducing significant thermal effects. In contrast, the outputs of the high-power phase modulators were connected to external 50-Ω terminators via the PCB connectors. This design ensured effective heat dissipation away from the chip, maintaining stable device operation under high RF drive power.

*Experimental setup*

Schematic diagrams of the experimental setups are shown in Figs. 2A, 3B, 4A and 5A of main text. A narrow-linewidth CW fiber laser (NKT E15, <100 Hz linewidth) served as the seed source for generating the EO comb (Fig. 2A). The modulators were driven by an arbitrary waveform generator (Keysight M8199A) with precise synchronization and phase delay adjustment between RF signals, enabling the synthesis of a flat-top and broad EO comb in the first modulation stage. The EO comb was then polarization-optimized by a polarization controller and amplified using an erbium-doped fiber amplifier (EDFA). Then, a fiber polarization controller was placed after the EDFA to ensure optimal polarization alignment in the second modulation stage. The second stage was driven by a microwave frequency comb generated by the same arbitrary waveform generator. EO combs 1 and 2 were then subsequently combined with an FML via an optical in-phase/quadrature demodulator (Optoplex, mixer, C-band) followed by a pair of balanced photodetectors (BPDs, Keyang, 10 GHz bandwidth) for heterodyne detection.

In the mode-hop FMCW ranging experiment (Fig. 4A), the FMCW ranging signal was also detected by an additional photodetector. Both the beat frequencies for absolute frequency calibration and the FMCW ranging signals were captured using a high-speed oscilloscope (R&S RTO6) with a sampling rate equaled to the repetition rate $f_{r,2}$.

In the FMCW LiDAR demonstration (Fig. 5A), a portion of the FML output was injected into port 1 of a fiber circulator. The light from the circulator's port 2 entered a fiber collimator and propagated through free space to illuminate the target. The reflected light was captured by the same collimator. Then, the reflected light traveled back through port 2 of the circulator and exited via port 3. A dual-axis galvanometer scanner (Thorlabs GVS012) steered the collimated beam over the scene, operating at linear scan rates of either 2 Hz or 0.5 Hz up to the imaging target.

**Acknowledgments:**

We thank Dr. Zhongwen Deng from Xidian University for providing a part of testing equipment for the ranging experiments.

**Funding:** This work received support from National Natural Science Foundation of China (62227822, 52327805, 62435009).

**Author contributions:**

    Conceptualization: G.H.W., Y.L., W.W.Y., X.Y.J.

    Experiment: W.W.Y., X.Y.J.

    Device fabrication and assembly: J.Y.W., X.L.C., W.W.Y., X.Y.J.

    Funding acquisition: G.H.W., Y.L.

    Supervision: G.H.W., Y.L., X.L.C.

    Writing – original draft: W.W.Y., Y.L.

    Writing – review & editing: G.H.W., X.L.C., X.Y.J., J.Y.W.

**Competing interests:** Authors declare that they have no competing interests.

**Data and materials availability:** The data that supports the findings of this study are available from the corresponding authors upon reasonable request.

**Supplementary Materials**

Materials and Methods

Supplementary Text S1 to S6

Figs. S1 to S6

Table S1

References 47–52

Movies S1 to S2

# Supplementary Materials for

## Electro-optic frequency comb-empowered precise measurement of the dynamic frequency of a laser

Weiwei Yang, Xingyu Jia, Jingyi Wang, Xinlun Cai, Yang Li, Guanhao Wu

Corresponding author: caixlun5@mail.sysu.edu.cn; liyang328@mail.sysu.edu.cn; guanhaowu@mail.tsinghua.edu.cn

**The PDF file includes:**

Supplementary Text S1 to S6
Table S1
Movies S1 to S2

**Supplementary Text**

Text S1. Interference between a frequency-modulated laser and an optical frequency comb

We consider the interference between a frequency-modulated laser (FML) and an optical frequency comb to analyze the time-resolved behavior of the beat note and extract the instantaneous frequency of the FML (Fig. S1A). The FML has a time-varying frequency $f(t)$ and an initial phase $\phi_{\text{FML},0}$, and the electric field $E_{\text{FML}}(t)$ can be expressed as

$$E_{\text{FML}}(t) = A_{\text{FML}} \exp\left[j\left(2\pi \int_0^t f(t')\mathrm{d}t' + \phi_{\text{FML},0}\right)\right], \tag{S1}$$

where $A_{\text{FML}}$ is the field amplitude, and the exponential term represents the accumulated optical phase of the FML during its frequency chirp.

An optical frequency comb, commonly generated through electro-optic modulators, mode-locked lasers, or microresonators-based methods, comprises a series of discrete, equally spaced spectral lines characterized by a repetition rate $f_r$ and a carrier-envelope offset frequency $f_{\text{ceo}}$. The electric field of the $n_{comb}$-th comb line $E_n(t)$ can be written as

$$E_n(t) = A_n \exp[j(2\pi f_n t + \phi_{n,0})], \tag{S2}$$

where $f_n = n_{comb} \times f_r + f_{\text{ceo}}$; here, $A_n$ denotes the amplitude, $f_n$ is the corresponding optical frequency, and $\phi_{n,0}$ is the initial phase of the $n_{comb}$-th comb line. According to the Nyquist sampling theorem, when the sampling rate $f_s$ is equal to the repetition rate $f_r$, only the beat note corresponding to the nearest comb line can be resolved. The heterodyne beat between the FML and the optical frequency comb is detected using an in-phase/quadrature (I/Q) optical demodulator. The real-valued signals from the heterodyne beat from the in-phase and quadrature detection channels are individually expressed as

$$\begin{aligned} s_{\text{I}}(t) &= \text{Re}[E_{\text{FML}}(t) \cdot E_n^*(t)] \\ s_{\text{Q}}(t) &= \text{Re}\left[e^{j\frac{\pi}{2}} E_{\text{FML}}(t) \cdot E_n^*(t)\right] = \text{Im}[E_{\text{FML}}(t) \cdot E_n^*(t)] \end{aligned}. \tag{S3}$$

Therefore, the complex envelope can be expressed as

$$s(t) = s_{\text{I}}(t) + j s_{\text{Q}}(t) = E_n^*(t) E_{\text{FML}}(t) = A(t) e^{j\theta(t)}, \tag{S4}$$

where $s(t)$ is the complex-valued IQ signal, $A(t)$ denotes the complex amplitude component of the beat signal, which is commonly excluded from the frequency extraction process, and $\theta(t)$ is the instantaneous phase defined by

$$\theta(t) = 2\pi\left(\int_0^t f(t')\mathrm{d}t' - f_n t\right) + \phi_{\text{FML},0} - \phi_{n,0}. \tag{S5}$$

*S1.1 Continuous-time instantaneous frequency extraction*

To extract the instantaneous frequency, we differentiate the beat phase with respect to time $\frac{d\theta(t)}{dt} = 2\pi f(t) - 2\pi f_n$. Consequently, the instantaneous beat frequency is given by $f_{\text{inst}}(t) = \frac{1}{2\pi}\frac{d\theta(t)}{dt} = f(t) - f_n$. This allows us to determine the absolute frequency of the FML as

$$f(t) = f_n + f_{\text{inst}}(t). \tag{S6}$$

*S1.2 Discrete-time instantaneous frequency extraction*

In the discrete-time domain, the heterodyne signal is sampled at uniform time intervals $\Delta T = 1/f_r$, yielding a discrete sequence $s_k = s(t_k)$, where $t_k = k\Delta T$ and $k \in \mathbb{Z}$. The phase of the heterodyne beat signal is first extracted as $\theta_k = \arg[s_k]$, then unwrapped to yield a continuous phase sequence $\theta_k^{\text{unwrap}}$. The instantaneous beat frequency is then calculated by numerically differentiating the phase sequence. A straightforward approach is to apply the forward difference method to express the instantaneous beat frequency $f_{\text{inst}}(t_k)$ as

$$f_{\text{inst}}(t_k) \approx \frac{1}{2\pi} \cdot \frac{\theta_{k+1}^{\text{unwrap}} - \theta_k^{\text{unwrap}}}{\Delta T}. \tag{S7}$$

This forward difference provides an approximate estimate of the beat frequency centered at $t_k + \frac{\Delta T}{2}$. However, it exhibits high sensitivity to phase noise and unwrapping errors.

A more robust and symmetric approach is to use the central difference method as

$$f_{\text{inst}}(t_k) \approx \frac{1}{2\pi} \cdot \frac{\theta_{k+1}^{\text{unwrap}} - \theta_{k-1}^{\text{unwrap}}}{2\Delta T}. \tag{S8}$$

This yields a phase-noise-tolerant estimate of the beat frequency centered at $t_k$. To ensure the validity of this assumption, the frequency of the FML must remain within the same comb mode spacing $f_r$ over the effective time window of the derivative (Figs. S1C, S1F). Hence, the maximum chirp rate $C$ could be expressed as:

$$C = \left|\frac{df(t)}{dt}\right| \leq \frac{f_r}{4\Delta T} = \frac{f_r^2}{4}. \tag{S9}$$

For the forward difference method, the maximum chirp rate limit should be $C \leq f_r^2/2$. Actually, this is a lower bound (Figs. S1B, S1E) related to the theory of sampling non-baseband signal[47].

For a linearly chirped FML, the phase excursion of the heterodyne beat can be described by the given equation $\Delta\phi = 2\int_{T_{\text{eff}}} C_{\text{chirp}}\, dt'$, where $T_{\text{eff}}$ is the effective interference time between the FML and the single nearest comb line, and $C_{\text{chirp}}$ denotes the constant chirp rate of the FML. As illustrated in Figs. S1E and S1F, under the limiting condition, a small accumulated phase $\Delta\phi$ resulting from a short interference time can significantly influence the accuracy of frequency measurement. In practice, the chirp rate is maintained well below the theoretical limit to ensure reliable and robust frequency extraction (Fig. S1D).

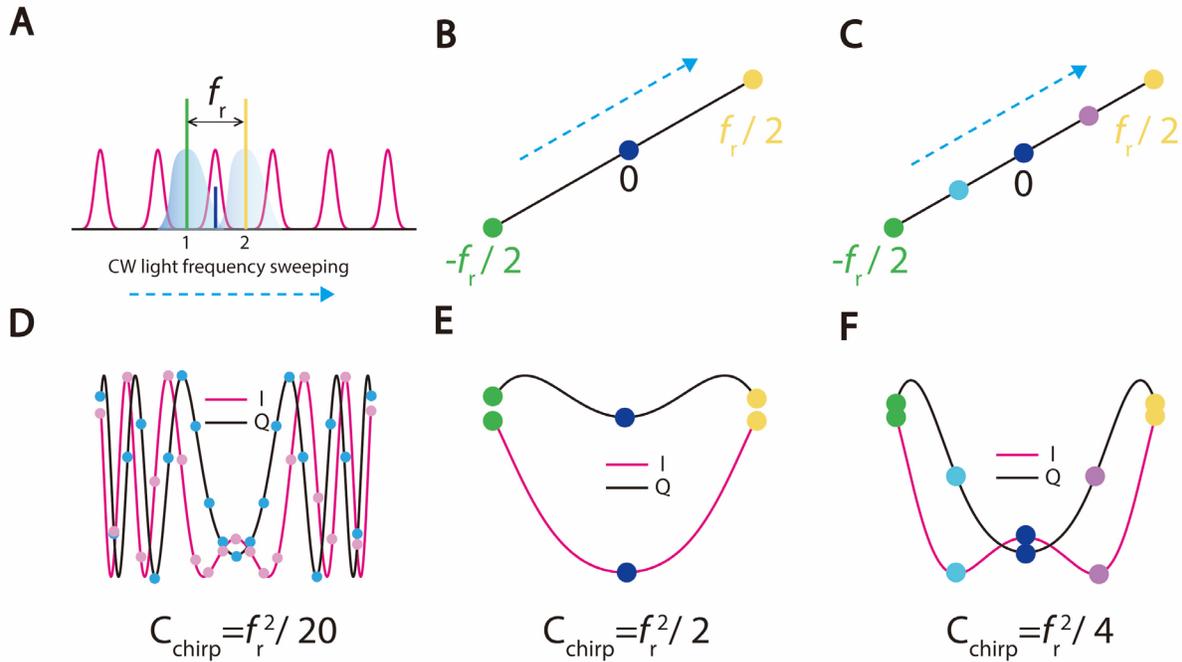

**Fig. S1. Schematic of the chirp rate limitation theory**. (**A**) Schematic of FML sweeps across a comb line under the time–bandwidth limit. (**B-C**) Time-varying behavior of beat notes. (**D-F**) Time-domain I and Q signals of the interference between the FML and the comb line. The colored circles represent valid sampling points.

Text S2. Broadband electro-optic frequency comb generation

Recently, various techniques have been developed for optical frequency comb generation[25], including mode-locked lasers (MLLs), Kerr frequency combs (Kerr combs), and electro-optic (EO) frequency combs. Distinct from MLLs and Kerr combs whose repetition rate tuning range is limited, waveguide-based EO combs offer exceptional flexibility in rapidly and widely tuning their repetition rates. However, waveguide-based EO combs still face limitations such as restricted spectral flatness, limited bandwidth, and the difficulty of realizing cascaded high-performance modulation stages on-chip.

To generate an EO frequency comb with a flat-top, broadband spectrum and an arbitrarily tunable repetition rate, we cascaded one amplitude modulator (AM) with three phase modulators (PMs) on a single thin film lithium niobate chip (Fig. S2A), followed by an AM driven by a microwave frequency comb.

*S2.1 Theoretical model and simulation result*

Consider a continuous-wave (CW) laser with an optical angular frequency $\omega_c$, and its optical field can be expressed as

$$E_{\text{in}}(t) = E_0 e^{j\omega_c t}, \tag{S10}$$

where $E_0$ is the amplitude of the optical field. A single EO phase modulator driven by a modulation signal $V(t) = V_{\text{PM}} \sin(\omega_m t)$ modulates the phase of the input optical field. Here, $V_{\text{PM}}$ is the amplitude, and $\omega_m$ is the angular frequency of the radio frequency (RF) modulation signal. Assuming the phase modulator has a half-wave voltage of $V_\pi^{\text{PM}}$, the output optical field $E_{\text{PM}}(t)$ in the time domain can be expressed as

$$E_{\text{PM}}(t) = E_0 e^{j\omega_c t} e^{j m_{\text{PM}} \sin(\omega_m t)}, \tag{S11}$$

where $m_{\text{PM}} = \pi \times V_{\text{PM}}/V_\pi^{\text{PM}}$ is the modulation index of the PM, which is mainly determined by the modulator design and the power of the applied modulation signal.

The frequency domain expression of the output optical field, obtained via the Fourier transform, can be written as

$$\widetilde{E_{\text{PM}}}(\omega) = 2\pi E_0 \sum_{n=-\infty}^{\infty} J_n(m_{\text{PM}}) \delta(\omega - n\omega_m - \omega_c). \tag{S12}$$

The term $J_n(m_{\text{PM}})$ is the Bessel function of the first kind of order, which determines the amplitude of the $n$-th comb line in the spectrum. The Dirac delta function $\delta(\omega - n\omega_m - \omega_c)$ indicates that the output spectrum consists of discrete comb lines located at $\omega_n = n\omega_m + \omega_c$. Equation (S12) highlights two main limitations of EO comb generated by a single PM. First, the comb spectrum suffers from poor flatness due to the nonuniform amplitude distribution of Bessel functions. Second, the number of comb lines is fundamentally limited by the modulation index, restricting the spectral bandwidth of the EO comb.

To improve the spectral flatness of the EO comb, we can cascade an amplitude modulator with one or more phase modulators. First, we consider the case of cascading a single PM. If modulation signals with the same angular frequency $\omega_m$ are applied to the modulators, the time-domain modulation signals can be expressed as $V_{\text{AM}}(t) = V_{\text{AM}} \sin(\omega_m t)$ and $V_{\text{PM}}(t) = V_{\text{PM}} \sin(\omega_m t + \phi)$, where $\phi$ represents the relative phase delay between the AM and PM drive signals.

Assuming the amplitude modulator has a half-wave voltage of $V_\pi^{\text{AM}}$ and is biased with a direct current (DC) voltage $V_{\text{DC}}$, the resulting optical output field $E_{\text{AM}}(t)$ can be expressed as

$$E_{\text{AM}}(t) = E_0 e^{j\omega_c t} \sum_{n=-\infty}^{\infty} \cos\left(\varphi_{\text{DC}} + \frac{n\pi}{2}\right) J_n(m_{\text{AM}}) e^{j n \omega_m t}. \tag{S13}$$

Here, the modulation index and DC-induced phase shift are defined as $m_{\text{AM}} = \pi \times V_{\text{AM}}/V_\pi^{\text{AM}}$ and $\varphi_{\text{DC}} = \pi \times V_{\text{DC}}/V_\pi^{\text{AM}}$, respectively. By tuning the DC bias, one can selectively enhance or suppress specific comb lines in the spectrum. The total output field after cascading the AM and PM can now be written as $E_{\text{out}}(t) = E_{\text{AM}}(t) e^{j m_{\text{PM}} \sin(\omega_m t + \phi)}$. Substituting the expression for $E_{\text{AM}}(t)$ and expanding the PM term, the output optical field can be expressed as

$$E_{\text{out}}(t) = E_0 e^{j\omega_c t} \sum_{n=-\infty}^{\infty} \sum_{k=-\infty}^{\infty} \cos\left(\varphi_{\text{DC}} + \frac{n\pi}{2}\right) J_n(m_{\text{AM}}) J_k(m_{\text{PM}}) e^{jk\phi} e^{j(n+k)\omega_m t}. \quad (S14)$$

Therefore, we obtain the frequency-domain representation of the output optical field as

$$\widetilde{E_{\text{out}}}(\omega) = 2\pi E_0 \sum_{N=-\infty}^{\infty} A_N(\phi) \delta(\omega - \omega_c - N\omega_m), \quad (S15)$$

where $N = n + k$ and the amplitude of the $N$-th comb line is given by

$$A_N(\phi) = \sum_{n=-\infty}^{\infty} \cos\left(\varphi_{\text{DC}} + \frac{n\pi}{2}\right) J_n(m_{\text{AM}}) J_k(m_{\text{PM}}) e^{j(N-n)\phi}. \quad (S16)$$

Accordingly, the output light field can be viewed as an EO comb comprising discrete comb lines located at $\omega_c + N\omega_m$. The amplitudes of the individual comb lines are determined by the convolution of the AM and PM Bessel-function weightings, with the relative phase $\phi$ serving as an adjustable parameter for spectral shaping.

While the use of a cascaded amplitude modulator can effectively flatten the spectrum of the EO comb, it does not address the inherent limitation in spectral bandwidth, which is still constrained by the limited microwave power that can be applied to individual phase modulators. The Bessel function $J_n(m_{\text{PM}})$ for large $n$ decays rapidly unless the modulation index $m_{\text{PM}}$ is large, making it difficult to generate high-order sidebands with a single phase modulator. Therefore, a wider spectral bandwidth requires a larger $m_{\text{PM}}$. On the one hand, the achievable modulation index is constrained by the maximum RF drive voltage and the half-wave voltage $V_\pi^{\text{PM}}$ of the modulator. Considering power consumption, the RF drive voltage $V_{\text{PM}}$ is usually limited, constraining the increase of the modulation index $m_{\text{PM}}$. On the other hand, for a fixed drive voltage $V_{\text{PM}}$, the modulation index $m_{\text{PM}}$ is inversely proportional to $V_\pi^{\text{PM}}$. Notably, the RF half-wave voltage $V_\pi^{\text{PM}}$ is influenced by the length of the modulator, which indicates that we can enhance $m_{\text{PM}}$ by increasing the modulation length. To realize a longer effective modulation length without increasing the device size, a folded modulator structure can be employed, effectively extending the modulation length and thereby reducing the RF half-wave voltage $V_\pi^{\text{PM}}$. A practical approach to further increase the number of comb lines is to cascade multiple phase modulators. By sequentially applying phase modulation stages, the overall modulation index $m_{\text{PM}}^{\text{all}}$ can be effectively increased while simultaneously reducing the high RF power requirement for each phase modulator.

Overall, by cascading an amplitude modulator with multiple phase modulators, one can generate a broadband and flat-top EO comb. The amplitude modulator improves the flatness of the EO comb. Meanwhile, each phase modulator adopts a folded waveguide design to increase the modulation index, thereby generating higher-order sidebands. Besides, additional phase modulators can be cascaded to further extend the spectral bandwidth.

Numerical simulation is implemented to optimize the EO comb generation, based on cascaded AM and PMs using the OptiSystem software. In our simulation, we obtained EO-comb spectra generated by one, two, and three phase modulators, all operating at a repetition rate of 25 GHz (Fig. S2B). As illustrated from top to bottom, the results show that increasing the number of phase modulators results in a progressively broader frequency comb with a larger number of comb lines.

### *S2.2 Design and fabrication of an integrated EO comb*

Based on the principle of generating a broadband and flat-top EO comb, we propose a scheme that fully leverages the excellent electro-optic modulation performance of thin-film lithium niobate. In this design, one amplitude modulator and three phase modulators are monolithically integrated on a single chip, allowing compact implementation and efficient modulation.

The optical signal enters the chip through the bottom edge coupler and sequentially passes through one AM and three PMs (Figs. S2A, S2D). The DC bias electrodes are placed on the top side of the chip (four metal squares in Fig. S2D), while the RF input pads for the AM and PMs are located on the left side, and the RF output pads of the PMs are positioned on the right side. To further reduce the device size while maintaining a low half-wave voltage, each phase modulator adopts a folded waveguide structure. As illustrated in Fig. S2E, the bending regions employ an air-bridge-assisted traveling-wave electrode (TWE) design to mitigate the degradation of high-frequency performance caused by bending in conventional TWE structures. As shown in Fig. S2D, within a compact chip area of $1\,\text{cm} \times 1\,\text{cm}$, the layout accommodates one folded amplitude modulator and three folded phase modulators. As depicted in Figure S2C, the packaged device only requires microwave drive signals and terminators, along with optical fiber pigtails, making it highly convenient for on-site experiments.

### *S2.3 The second stage denser EO comb generation*

A higher repetition rate $f_{r,1} = \omega_m/2\pi$ facilitates the generation of broader bandwidth EO combs by increasing the frequency spacing between adjacent comb lines. However, a higher repetition rate also increases the demand on the data acquisition system, posing challenges for system implementation. To maintain a broad spectral bandwidth while reducing the repetition rate, we introduced a secondary modulation stage for EO-comb generation. Specifically, we used an amplitude modulator driven by a microwave frequency comb with a repetition rate $f_{r,2}$ to modulate the primary EO comb, resulting in a secondary EO comb with a reduced repetition rate of $f_{r,2}$.

We consider the generation of a secondary EO frequency comb (also referred to as a sub-comb) from a single comb line of the primary EO comb, with angular frequency $\omega_{c1}$ (i.e., frequency $f_{c1} = \omega_{c1}/2\pi$). This optical carrier is modulated by a microwave frequency comb

$$V_{\text{comb}}(t) = \sum_{q=1}^{Q} V_q \cos(2\pi n f_{r,2} t + \phi_q). \tag{S17}$$

where $V_q$ and $\phi_q$ represent the amplitude and phase of the $q$-th microwave comb line, respectively. The maximum harmonic index $Q$ is limited by the bandwidth of the RF signal source.

To ensure accurate and predictable sideband generation, the secondary electro-optic comb generation is operated within the small-signal linear regime of the amplitude modulator. In this regime, the modulator's output optical field can be assumed linearly proportional to the driving voltage. This assumption simplifies the analysis and enables direct mapping from the microwave frequency comb to the generated optical sidebands. The amplitude modulator output field under the linear approximation can then be written as

$$E_{\text{out}}^{(2)}(t) \approx \frac{\pi}{V_\pi} V_{\text{comb}}(t) \cdot e^{j\omega_{c1} t}. \tag{S18}$$

By substituting $V_{\text{comb}}(t)$ into the above expression, we obtain

$$E_{\text{out}}^{(2)}(t) = \frac{\pi}{V_\pi} \sum_{q=1}^{Q} V_q \cos(2\pi q f_{r,2} t + \phi_q) e^{j\omega_{c1} t}, \tag{S19}$$

which can be simplified as

$$E_{\text{out}}^{(2)}(t) = \frac{\pi}{2V_\pi} \sum_{q=1}^{Q} V_q \left[ e^{j(\omega_{c1} + 2\pi q f_{r,2})t + j\phi_q} + e^{j(\omega_{c1} - 2\pi q f_{r,2})t - j\phi_q} \right]. \tag{S20}$$

Taking the Fourier transform of $E_{\text{out}}^{(2)}(t)$, we obtain discrete comb lines located at $f_{\text{EO}} = f_{c1} \pm q f_{r,2}$, $q = 1, 2, \cdots, Q$. Our method not only maintains the broadband spectrum of the EO comb but also allows arbitrary adjustment of the repetition rate, providing a flexible approach for applications that require both a broad optical bandwidth and a tunable repetition rate, such as optical communications[48], high-resolution spectrometer[49], and repetition rate modulated frequency comb LiDARs[50].

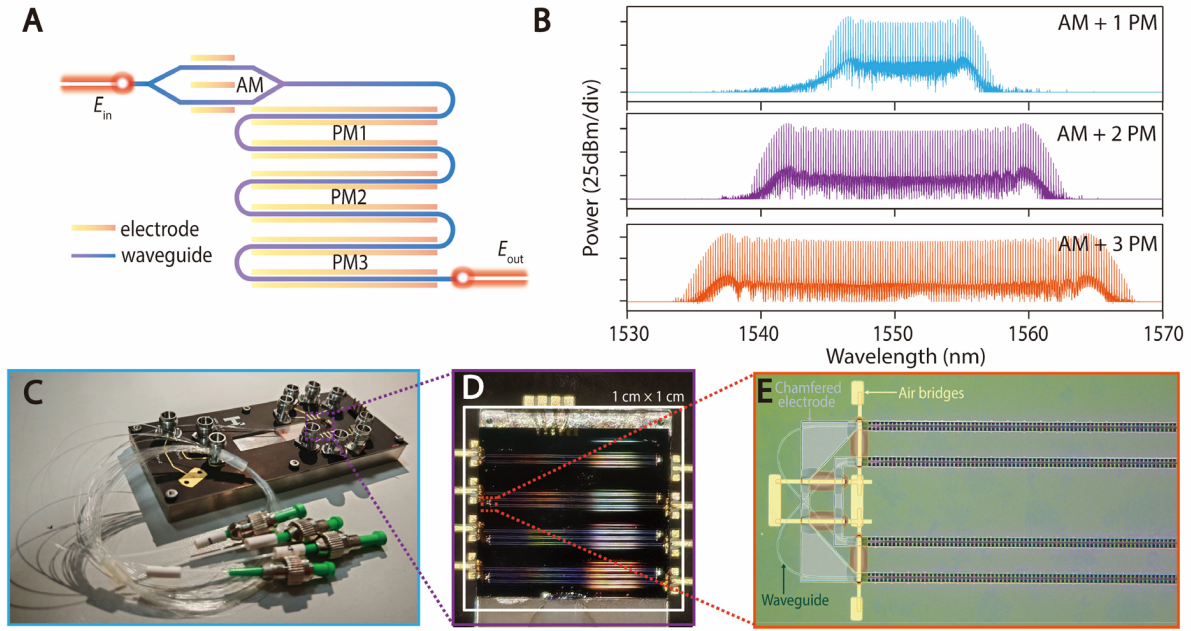

**Fig. S2. Generation of highly tunable flat-top EO comb**. (**A**) Schematic of cascaded amplitude and phase modulators. (**B**) Simulated EO-comb spectra generated by one, two, and three phase modulators (top to bottom) with a 25-GHz repetition rate. (**C**) Packaged device photograph. (**D**) Photograph of the TFLN chip. (**E**) Microscope image of the folded region of the fabricated modulator.

Text S3. Generation of arbitrarily frequency-modulated lasers

To achieve precise microwave-to-optical frequency mapping, we utilize an electro-optic modulation scheme driven by an arbitrary waveform generator (AWG). A frequency-stabilized CW laser is employed as the seed source and injected into an electro-optic amplitude modulator. The modulator is driven by the AWG that delivers a programmable microwave signal—such as the "THU" waveform illustrated in Fig. S3. Through electro-optic modulation, the frequency characteristic of the microwave driving signal is directly transferred to the optical domain, resulting in the generation of a repetition rate-modulated frequency comb.

To select a specific comb line from the repetition rate-modulated frequency comb with high resolution, an optical filtering module with a fiber Bragg grating (FBG) filter and an optical circulator is implemented (Fig. S3). The repetition rate-modulated frequency comb is injected into port 1 of the optical circulator and routed to port 2, where it enters the FBG filter. The FBG filter reflects only the designated comb line corresponding to its center wavelength, while transmitting the remaining spectral components. The reflected comb line is routed back to port 2 and subsequently directed to port 3 of the circulator. This design enables arbitrary encoding and decoding between microwave and optical signals.

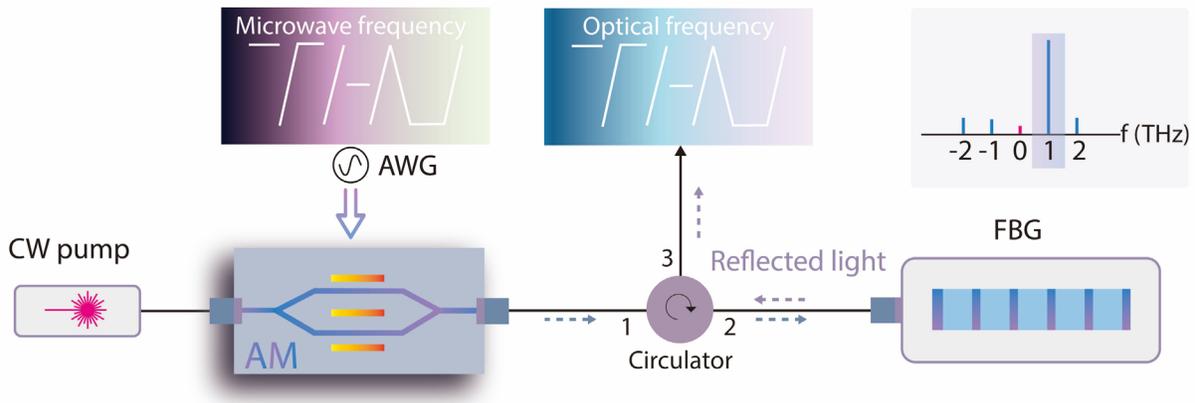

**Fig. S3. Generation of an arbitrarily frequency-modulated laser driven by a microwave signal.** AWG: arbitrary waveform generator. FBG: fiber Bragg grating.

Text S4. Evaluation of absolute frequency measurement performance

We validate the system's frequency measurement uncertainty by measuring the frequency of a stabilized CW laser (Fig. S4A). Within an ultrashort measurement window of approximately 1 ns, the frequency uncertainty is measured to be around 6.5 MHz, mainly limited by phase noise and the signal-to-noise ratio (SNR) of the detection system. Extending the averaging time leads to improvement in precision, resulting in an uncertainty of approximately 1.3 kHz over a 10-μs averaging interval. This result demonstrates the system's capability to track fast frequency excursions while maintaining high accuracy over longer averaging times.

Using the method described in Supplementary Text S3, we generated an FML with frequency steps of 20 MHz to test the system's resolution (Fig. S4B). Each step was held for 0.2 μs, demonstrating the system's ability to clearly distinguish these frequency changes. Figure S4B (top) shows the result obtained using the short-time Fourier transform (STFT), which applies a moving window for time–frequency analysis. However, the fixed window size imposes a trade-off between time and frequency resolution, resulting in limited resolution in the time-frequency representation[51]. By applying the Pseudo-Wigner–Ville distribution[52], we obtained a clearer time-frequency representation (Fig. S4B, bottom). Although the Pseudo-Wigner–Ville distribution ideally provides infinite resolution in both time and frequency domains due to the absence of averaging over finite intervals, the lack of clarity in the time-frequency plot is mainly caused by remaining cross-term interference inherent to the algorithm. Despite the presence of cross-term interference, the experimental results in Fig. S4B show that the system still achieves a frequency resolution better than 20 MHz.

To verify the system's adaptability to arbitrarily chirping and mode-hop behaviors, we conducted experiments under designed non-linear and discontinuous FML tuning conditions. Sinusoidal drive signals offer significant advantages for high-bandwidth frequency modulation. In this work, we employed our absolute frequency measurement method to characterize a sinusoidal drive with a 4-GHz modulation bandwidth and 20-ns modulation period (Fig. S4C), demonstrating the potential of characterizing high-speed sinusoidal chirps for advanced modulation schemes. Additionally, we generated sinusoidal signals with peak-to-peak amplitudes from 40 MHz to

160 MHz and periods between 0.25 µs and 0.125 µs using an AWG, and rapidly switched between these signals. Our absolute frequency measurement method accurately tracked frequency changes during rapid transitions (Fig. S4D), demonstrating robust and precise monitoring under dynamic conditions.

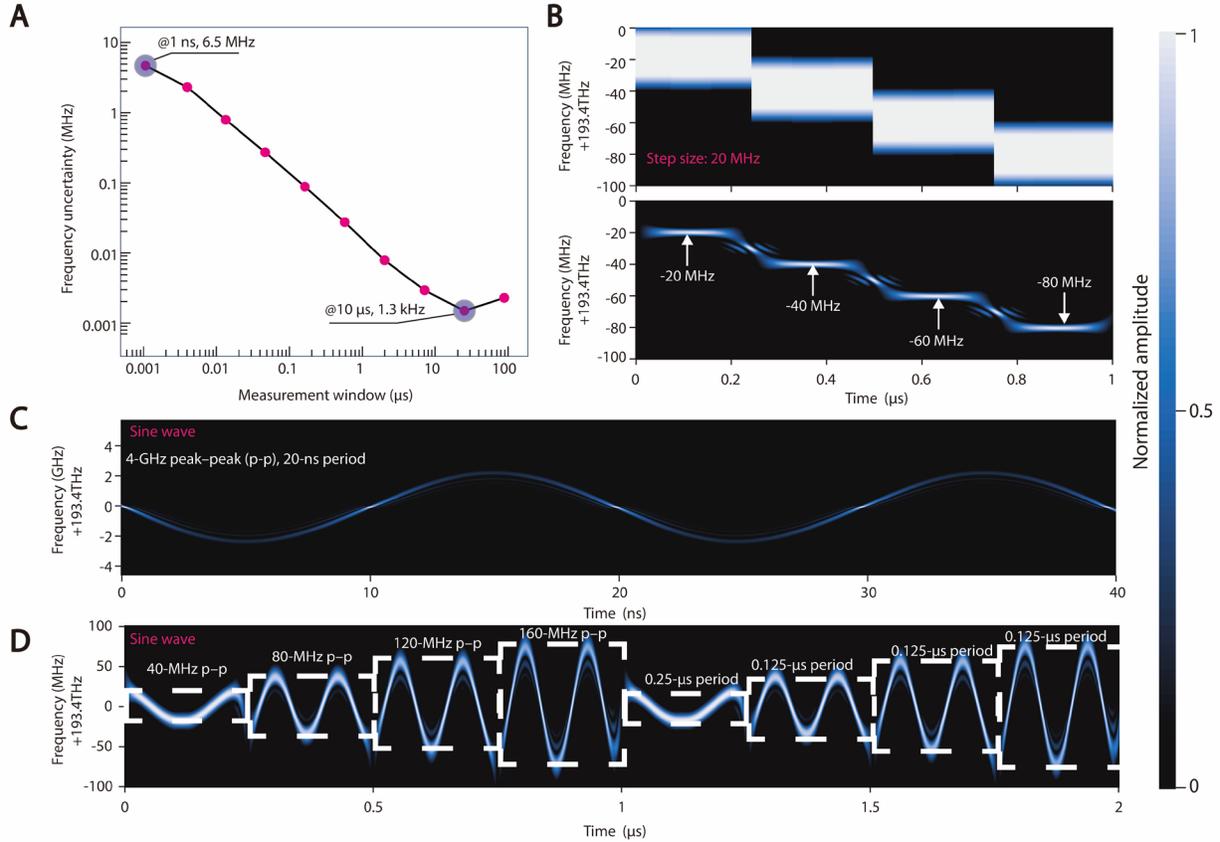

**Fig. S4. Schematic of frequency measurement uncertainty, resolution, and adaptability to arbitrarily chirping and mode-hop behaviors**. (**A**) Frequency measurement uncertainty as a function of measurement window. (**B**) Demonstration of accurate frequency tracking for a 20-MHz step-hop FML. (**C**) Absolute frequency measurement of a sinusoidally chirped FML (20-ns period, 4-GHz peak-to-peak). (**D**) Absolute frequency measurement of an FML with concatenated sinusoidal chirps of varying amplitudes and periods.

Text S5. Reconstruction of nonlinear, mode-hop FMCW ranging signals

This section derives a theoretical model for distance extraction from the interference signal in a conventional Mach–Zehnder FMCW system (Fig. S5A). The derived model serves as the foundation for further analysis under nonlinear chirping and mode-hop conditions. Within the interferometer, the chirped light is divided into two paths: a reference arm and a measurement arm. The detected electric field is composed of the reference signal $E_{\text{ref}}(t) = E_{\text{FML}}(t)$ and the delayed measurement signal $E_{\text{sig}}(t) = E_{\text{FML}}(t - \tau)$, where $\tau = D/c$ represents the time delay, with $D$ representing the optical path difference (OPD) between the two arms and $c$ is the speed of light in vacuum. The interference signal $I(t)$ can be expressed as

$$I(t) \propto \text{Re}[E_{\text{ref}}(t) E_{\text{sig}}^*(t)] = A_{\text{FML}}^2 \cos[\Phi(t)], \tag{S21}$$

where the instantaneous phase $\Phi(t)$ is

$$\Phi(t) = 2\pi \left( \int_0^t f(t')dt' - \int_0^{t-\tau} f(t') \, dt' \right) = 2\pi \int_{t-\tau}^t f(t')dt'. \tag{S22}$$

Because the delay $\tau$ is typically very small, we perform the first-order Taylor expansion of the instantaneous frequency $f(t')$ around the point $t$, the instantaneous phase can be approximated as $\Phi(t) \approx 2\pi f(t)\tau$. Accordingly, the interferometric signal can be approximated as

$$I(t) = A \cos[2\pi f(t)\tau], \tag{S23}$$

Here, $A$ is an amplitude modulation term. The beat frequency observed in this signal is $f_b(t) = \frac{1}{2\pi} \frac{d\Phi(t)}{dt} \approx \tau \frac{df(t)}{dt}$. For the linear chirp case where $\beta = \frac{df(t)}{dt}$ is constant, the interference signal $I(t)$ is a single-frequency cosine wave under uniform time sampling, the Fast Fourier Transform (FFT) can be directly applied to accurately extract the beat frequency $f_b$. Subsequently, the OPD can be calculated as

$$D = \frac{c f_b}{\beta}, \tag{S24}$$

*S5.1 Nonlinear chirp and mode hop effects*

In practice, the instantaneous frequency $f(t)$ deviates from ideal linear behavior due to nonlinear tuning and mode hop, i.e., $f(t)$ is a general nonlinear function (Fig. S5D). The instantaneous beat frequency becomes time-dependent:

$$f_b(t) = \frac{1}{2\pi} \frac{d\Phi(t)}{dt} = f(t) - f(t-\tau) \neq \text{constant}. \tag{S25}$$

Thus, the interference signal $I(t)$ is a nonlinear chirp (Fig. S5B), leading to failure or errors in FFT-based distance extraction (Fig. S5E, black line).

*S5.2 Frequency-domain resampling and Lomb–Scargle periodogram*

To address nonlinear chirping and mode hop, we leverage prior knowledge of the instantaneous optical frequency of the FML, obtained from the absolute frequency measurement module. Because the instantaneous frequency sequence $f_l = f(t_l)$ is directly available, we plot the measured signal as a function of $I(f_l)$ (Fig. S5C). Equation (S23) can be expressed as a single-frequency cosine waveform

$$I(f_l) = A \cos(2\pi f_l \tau). \tag{S26}$$

This formulation transforms the nonlinear time-domain interference signal into a sinusoidal signal sampled over the optical frequency axis $f_l$. However, this representation breaks the uniform time-domain sampling structure, preventing the application of classical FFT (Fig. S5C). Unlike FFT, which requires uniform time sampling, the Lomb-Scargle periodogram is suitable for non-uniform time sampling. The Lomb-Scargle periodogram can be viewed as fitting a sinusoidal model to unevenly sampled data.

To estimate $\tau$, we fit the interference signal to the following model

$$I_l = y_0 + A_T \cos(2\pi f \times f_l + \phi_T) + \epsilon_l, \tag{S27}$$

Where $f$ is a trial frequency which corresponds to the delay. Here, $I_l$ denotes the measured signal at the $l$-th frequency sample $f_l$. The term $y_0$ represents a floating mean or DC offset that accounts for the baseline of the signal. $A_T$ is the amplitude of the sinusoidal component, and $\phi_T$ is the phase offset. Finally, $\epsilon_l$ denotes additive noise or measurement error at the $l$-th sample. The Lomb-Scargle periodogram inherently removes the effect of $y_0$ by centering the data, and assumes $\epsilon_l$ to be zero-mean and uncorrelated noise, so neither $y_0$ nor $\epsilon_l$ significantly influences the estimation of $\tau$. The normalized Lomb–Scargle power spectrum can be expressed as

$$P(f) = \frac{1}{2\sigma^2} \left( \frac{[\sum_l (I_l - \bar{I}) \cos(2\pi f(f_l - \phi_T))]^2}{\sum_l \cos^2(2\pi f(f_l - \phi_T))} + \frac{[\sum_l (I_l - \bar{I}) \sin(2\pi f(f_l - \phi_T))]^2}{\sum_l \sin^2(2\pi f(f_l - \phi_T))} \right), \tag{S28}$$

with the phase shift $\phi_T$ defined as $\tan(4\pi f \phi_T) = \sum_l \sin(4\pi f \times f_l) / \sum_l \cos(4\pi f \times f_l)$ and $\bar{I} = (1/L) \sum_{l=1}^{L} I_l$, $\sigma^2 = (1/(L-1)) \sum_{l=1}^{L} (I_l - \bar{I})^2$. After finding the peak frequency $f_{\text{peak}}$ in the Lomb–Scargle power spectrum (Fig. S5E, red line), the OPD can be calculated by

$$D = c \times f_{\text{peak}}. \tag{S29}$$

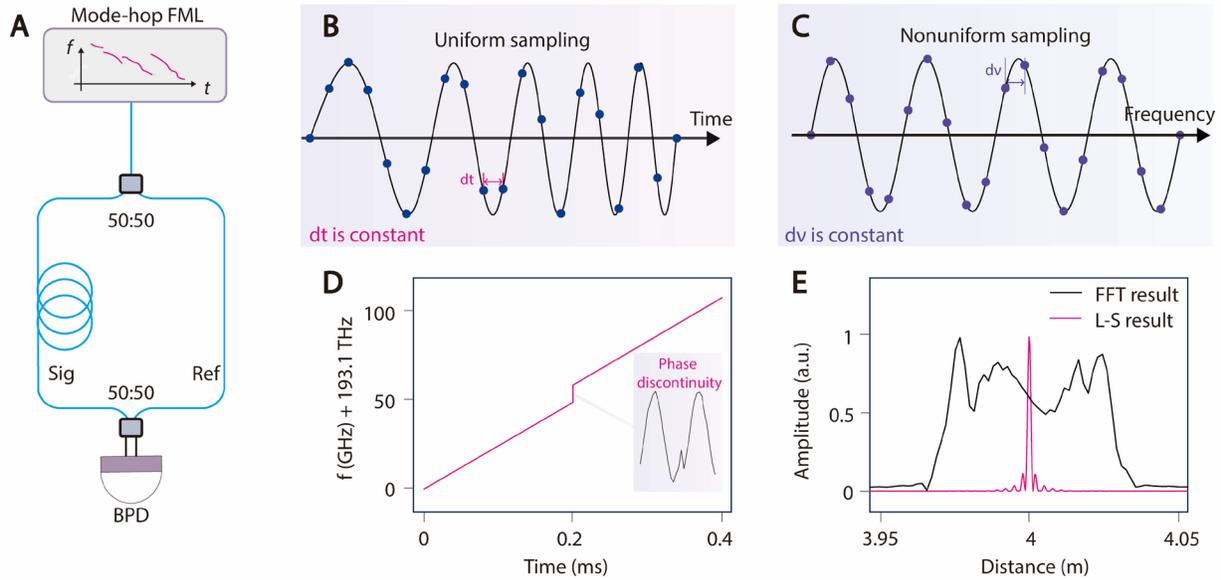

**Fig. S5. Schematics and simulations of reconstruction algorithm for FMCW ranging using mode-hop, nonlinear-chirping FML**. (**A**) Schematic of the Mach–Zehnder FMCW system, BPD: balanced photodetector. (**B**) Sampled uniformly in the time domain, (**C**) resampled uniformly in the optical frequency domain. (**D**) Simulated nonlinear, mode-hop FML for absolute frequency measurement, illustrating the phase discontinuity in the FMCW ranging signal at the mode-hop point. (**E**) Distance estimation using direct FFT (black line) vs. Lomb–Scargle (L-S) periodogram (red line).

Text S6. Performance evaluation of commercial mode-hop lasers for FMCW ranging

To further validate the effectiveness of the absolute frequency measurement module in the measurement of unknown frequency of FMLs, we characterized the mode-hop behavior of a commercial distributed feedback (DFB) laser and demonstrated its application in FMCW-based 3D imaging. Fig. S6A illustrates the frequency tuning behavior of the commercial DFB laser. Within a 1-ms sweep period, the laser covers a total tuning bandwidth of 25 GHz. A bidirectional overlapping sweep occurs during the first 0.35 ms (the highlighted area), with a mode-hop around 0.5 ms and severe nonlinearity throughout the sweep.

Distance spectrum calibration is enabled by using the measured absolute frequency. The original ranging signal (Fig. S6B) was reconstructed following the algorithm detailed in Supplementary Text S5, producing a frequency-domain reconstructed signal (Fig. S6D). The reconstructed signal effectively addresses the challenges of the non-linear sweep characteristics, such as bidirectional overlapping, severe nonlinearity, and mode-hop. Distance spectrum analysis was conducted on both calibrated and uncalibrated FMCW ranging signals. Before calibration, the distance peak is ambiguous (Fig. S6C, black line), making it difficult to identify the distance of the target. After applying the calibration, the peak becomes unambiguous, enabling effective distance resolving.

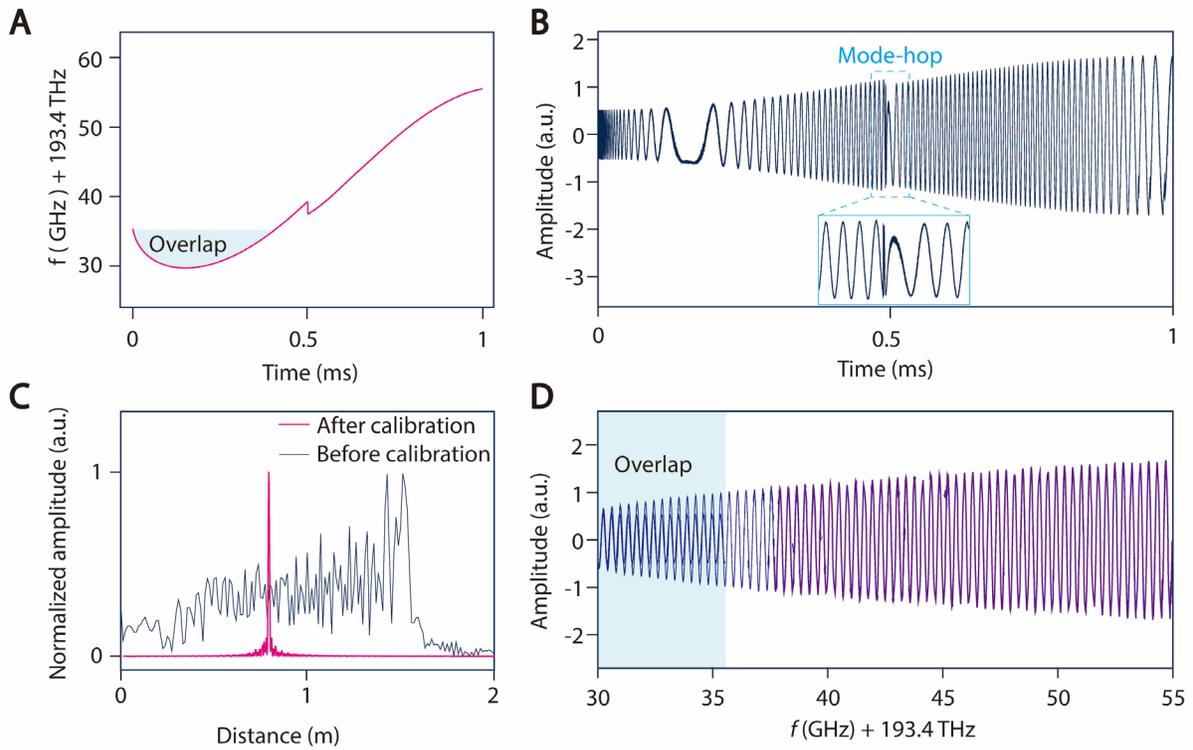

**Fig. S6. Reconstruction of ranging signals from a commercial DFB laser with mode hop, nonlinear chirping, and spectral overlap**. (**A**) Absolute frequency measured using our method. (**B**) FMCW ranging signal without calibration. (**C**) Distance estimation by using direct FFT (black) to signal in (**B**) vs. Lomb–Scargle analysis after optical-frequency resampling (red) in (**D**). (**D**) FMCW ranging signal resampled in the optical frequency domain.

**Table S1. Detailed performance comparison with other works**

| Method | Cavity | Mode-locked laser | | | | Kerr comb | | | EO comb |
|---|---|---|---|---|---|---|---|---|---|
| Absolute frequency measurement | **Yes** | No | **Yes** | **Yes** | No | No | **Yes** | No | **Yes** |
| Chirp rate(THz/s) | 1 | 1 | 1500 | 20 | 1 | 12.5 | 10 | 320,000 | **2,000,000** |
| Size of system | Large | Large | Large | Large | Large | **Chip-scale** | **Chip-scale** | **Chip-scale** | **Chip-scale** |
| Continuous frequency monitoring | No | No | Yes | **Yes** | No | No | **Yes** | No | **Yes** |
| Repetition rate(MHz) | \ | 250 | 100 | 100 | 250 | 48970 | 22000 | 50080 | 200~10000 |
| Acquisition rate(MHz) | 5 | ~10 | 100 | \ | \ | 2.5 | 5000 | 48 | 200~10000 |
| Frequency uncertainty | **~Hz** | \ | ~2kHz | ~kHz | \ | \ | 4MHz | \ | 1.3kHz |
| Article | Ref. (*12*[a]) | Ref. (*22*) | Ref. (*14*) | Ref. (*15*) | Ref. (*17*) | Ref. (*18*) | Ref. (*21*) | Ref. (*16*) | This work |

[a] Measuring the free spectral range of a fiber cavity with an electro-optic modulator.

**Movie S1 and Movie S2.**
Videos 1 and 2 present detailed 3D point clouds from varying perspectives.